\documentclass[twocolumn,showpacs,preprintnumbers,amsmath,amssymb]{revtex4}
\usepackage{graphicx}% Include figure files
\usepackage{dcolumn}% Align table columns on decimal point
\usepackage{bm}% bold math

\newcommand{\be}{\begin{equation}} \newcommand{\ee}{\end{equation}}
\newcommand{\bc}{\begin{center}} \newcommand{\ec}{\end{center}}
\newcommand{\bea}{\begin{eqnarray}} \newcommand{\eea}{\end{eqnarray}}
\newcommand{\tx}{\textstyle}
\begin{document}
\preprint{FT-031}

\title{Gauge invariance, radiative interferences and properties of vector mesons.\\}
% Force linebreaks with \\

\author{G. Toledo S\'anchez}

\affiliation{ Instituto de F\'{\i}sica, UNAM
A.~P.~20-364, M\'exico 01000 D.~F.~M\'exico\\}

\date{\today}% It is always \today, today,

\begin{abstract}
We state the implications on the properties of vector mesons due to gauge invariance. In particular, we find that polarized vector mesons exhibit a property in the radiation distribution of order $\omega^{-1}$ in the photon energy, namely  it is null when the gyromagnetic ratio becomes $g=2$. Therefore, the generalization of  the Burnett-Kroll theorem for polarized vector-meson states is held only if $g=2$. In addition, radiative interferences between the electric charge and any gauge invariant term is found to be parametrized by a common global factor which can be used to extract information of the involved states by a proper choice of the kinematical region, where they could be relevant.

\end{abstract}


\maketitle

\section{I. INTRODUCTION}

Vector mesons (states with quantum numbers $J^p=1^-$) and in general hadronic resonances are the kind of states that still offering challenges in particle physics. Although their quark composite nature in general fits within the quark model of the strong interaction, the lack of a reliable tool to handle QCD in the non-perturbative regime makes the predictions of their properties, by phenomenological models, a way to extract the most relevant ingredients of QCD. 
Experimental difficulties are also faced when such properties are tried to be measured. In particular, electromagnetic multipoles such as the Magnetic dipole moment (MDM) and the electric quadrupole moment have not been determined \cite{pdg} and then many models \cite{mackellar} are waiting to be on trial.\\
Elementary particles with similar properties are the weak interaction gauge bosons. The $W$ boson MDM is predicted  by the standard model to have ${\bf g}=2$ (here we will refer to the MDM by the gyromagnetic ratio ${\bf g}$ ).\\
Jackiw \cite{jackiw}, Ferrara \cite{ferrara}, Bargman \cite{bargman} and others have shown that this value enjoys many interesting features in the description of electromagnetic phenomena like the existence of a non electromagnetic gauge invariance, etc.\\
For vector mesons considered as gauge bosons of a hidden symmetry M. Bando \cite{bando} showed that  the coupling to an electromagnetic field  have  ${ \bf g}=2$ in a similar way to the $W$.
Then this particular value turns out to show up as one with implications in many aspects of the description of vector mesons.
In this work (chapter II) we offer an additional feature to favor it by observing the {\em radiative decay interferences for polarized vector mesons, there the  ${\cal O}(\omega^{-1})$ terms, in the photon energy ($\omega$), are null only if  ${\bf g =2}$}. Such condition is requested if we wish to keep the analogy to the non-polarized case, where it holds as stated by Burnett and Kroll \cite{burnett}.\\
The question of how can we measure the MDM of vector mesons is a problem that we have addressed and pointed out to the radiative decays as the most promising scenarios \cite{toledo9700}. The main idea is that the photon emission off the vector mesons carries valuable information on its electromagnetic structure and then some observables could show important dependence on such parameters. at the amplitude level  the leading contribution is the electric charge radiation and then the MDM radiation contribution. Turns out that they contribute in different kinematical regions of the photon energy spectrum, opening a possibility to extract the MDM. There, it is required to have under control  all the possible contributions coming from other sources, that we call model dependent. In chapter III we  show that the interference of the electric charge radiation with any gauge invariant term of the transition amplitude exhibit a typical structure, namely  one that allows to suppress such model dependent contribution, and then allowing to obtain the MDM. All the way around, we can enhance effects of intermediate states (model) to extract its relevant parameters.\\

\section{II. ${\bf g =2}$ FOR VECTOR MESONS}

Let us first make  some general remarks to clear the procedure followed along the paper.\\
Consider the decay amplitude for the process $i \rightarrow f+\gamma$. It  can be expressed as a power expansion in the photon energy $\omega$:

\be
{\cal M} =\frac{A}{\omega} +B\omega^0 + C\omega +\cdots 
\ee

The electric charges contribute at ${\cal O}(\omega^{-1})$ while the MDM enters at ${\cal O}(\omega^0$). Low's theorem \cite{low} guarantee that the first two terms are model-independent in the sense that they only depend on the non radiative decay amplitude and the multipoles involved.
The terms of higher order in $\omega$ involve model-dependent contributions (typically reflected by different intermediate states), and higher order electromagnetic multipoles. The squared amplitude, after summation over polarizations is:

\be
\sum_{pols \ i, f} |{\cal M}|^2 =  \frac{\alpha_2}{\omega^2}+ \alpha_0
\omega^0 + O(\omega)\ ,
\ee

The interference of ${\cal O}(\omega^{-1})$ vanishes. This is the Burnett-Kroll's theorem \cite{burnett}: The interference of the first and second terms of the amplitude expansion, in powers of the photon energy  $\omega$, {\it after sum over polarizations}, is null. Then  we can  naturally wonder about the polarized amplitudes. Do they also  show this vanishing feature? In the following we will show under which conditions it happens, namely when ${\bf g =2}$, taking the radiative decay of  the $\rho$ meson into two pions as a definite example.\\

{\it Polarized radiation distribution}. Lets now consider the decay $\rho^+ \rightarrow\pi^+  \pi^0  \gamma$, where the momenta is assigned as follows 
  $q$ $\rightarrow $ $p$ $p'$ $k$, $\epsilon$ and $\eta$ are the polarization 4-vector of the photon and vector meson respectively. The role of the MDM comes into the game through the $\rho\rho\gamma$ vertex, which can be written as:
\be
\Gamma^{\alpha\beta\mu} =  g^{\alpha\beta} (p+p')^\mu \alpha + \beta (k^\alpha g^{\mu\beta}-k^\beta g^{\mu\alpha})- \gamma(p+p')^\mu  k^\beta k^\alpha . \label{von}
\ee

The  form factors $\alpha$, $\beta$ and $\gamma$ depend on $k^2$ and at $k^2=0$ are related to the electric charge $q$ (in units of $e$), the magnetic dipole moment $\mu$ (in units of $e/2m_\rho$) and the electric quadrupole moment $\cal{Q}$ (in units of $e/m_\rho^2$) through the relations $\alpha=q=1$, $\beta=\mu$ and $\gamma=(1-\mu-{\cal Q})/2m_\rho^2$ \cite{nieves}. The gyromagnetic ratio ${\bf g}$ i.e. the magnitude of the magnetic dipole moment in such units, takes the canonical value  ${\bf g}=2$ in the case of elementary particles as the gauge boson $W$. However, besides the electric charge form factor which is fixed by electric charge conservation, no {\it ab initio} prescription exist for the form factors of non elementary particles,  like  hadronic resonances. Upon the identification of the electromagnetic nature of the coefficients in the single Feynman diagrams contributing to this process, it is possible to split the total decay amplitude in the form as described by Eq. (1).\\

Considering the soft photon  approximation, i.e. the first  two terms in Eq. (1), the vector meson electromagnetic vertex  is described up to the MDM contribution. In what follows we will compute the polarized amplitudes in the restframe of the decaying particle and the radiation gauge ($\epsilon_0=0$). A base for the vector meson polarization tensor is described by 3 vectors which are taken in a general form as $ \eta^{(j)}_\mu=(0,\vec{\eta}^{(j)}) $ with $j$ = 1, 2, 3, where 
\be
\vec\eta^{(1)}=\frac{1}{\sqrt{2}}(1,i,0)\hspace{.12cm}
\vec\eta^{(2)}=\frac{1}{\sqrt{2}}(1,-i,0)\hspace{.12cm} 
\vec\eta^{(3)}=(0,0,1).\nonumber
\ee

\noindent Using the configuration where $k=(\omega,0,0,\omega)$, in the gauge radiation, the transversality condition implies that $\vec k\cdot\vec\epsilon = \omega \epsilon_3 = 0$ and therefore the photon polarization tensor $\epsilon_\mu$ can be written as:

\[
\epsilon_\mu = (0,\epsilon_1 ,\epsilon_2 ,0).
\]
 
The polarized amplitudes under these conditions, denoted by ${\cal M}^{(1)}$, ${\cal M}^{(2)}$ and ${\cal M}^{(3)}$, depends explicitly on the gyromagnetic ratio as the only free form factor value. The 2 Lorentz  invariants which are needed to describe the amplitudes are chosen  be $p\cdot  k$ and $q\cdot k$ to exhibit the dependence on the photon energy. Then, the total transition probability for each of the 3 directions of the polarization becomes (except for a global factor $ (eg_{\rho\pi\pi})^2$):

\bea
\mid {\cal M}^{(1)}\mid^2 
=&\left(
\frac{\tx p\cdot\epsilon^*}{\tx p\cdot  k}
\right)^2 &
\left[ 2m^2_{\rho}\frac{p\cdot k}{q\cdot k}
\left(
N-\frac{p\cdot k}{q\cdot k}\right) -2m^2_\pi \right.\nonumber\\
&&\left. - 2p\cdot k \Delta
\left( N-2\frac{p\cdot k}{q\cdot k}\right) 
\right]\nonumber\\
&-\frac{\tx \epsilon^* \cdot \epsilon}{2} &
\left[{\bf g}\frac{p\cdot k}{q\cdot k}-2+N \Delta
\right]^2,\nonumber
\eea

\be
\mid {\cal M}^{(2)}\mid^2 = \mid {\cal M}^{(1)}\mid^2 
\ee

\bea
\mid {\cal M}^{(3)}\mid^2 =  \left(\frac{\tx p\cdot\epsilon^*}{\tx p\cdot k} \right)^2 &\left[
m^2_{\rho}\left( N-2\frac{p\cdot k}{q\cdot k}\right)^2
+ 4\left(\frac{p\cdot k}{m_{\rho}}\right)^2 \Delta^2 \right.\nonumber\\
&\left. +4 p\cdot k\Delta
 \left(N-2\frac{p\cdot k}{q\cdot k}\right)
\right].\nonumber
\eea

$\Delta \equiv 1-\frac{\bf g}{2}$, $N \equiv 1+\frac{(m^2_\pi -m^2_{\pi^0})^2}{m^2_{\rho}}$. A look to the terms  of  ${\cal O}(\omega^{-1})$ (by just counting the powers on $k$) reveals that all they are.proportional to $\Delta$. These terms are  then the interferences equivalent to the Burnett-Kroll ones for the non polarized case, and are not null, unless the magnetic dipole moment  takes the canonical value ${\bf g}=2$ \cite{toledoPRD02}. Thus this result offers a new feature of the effect of the MDM in polarized radiative decays, which favors a particular value to exhibit a vanishing of the  ${\cal O}(\omega^{-1})$ terms. It is worth to mention that in the non-polarized case the same value can lead to a vanishing of the total amplitude in a particular kinematical region \cite{samuel}. Adding the  three  polarized equations we render to the unpolarized case:

{\samepage
\begin{eqnarray}
 \sum_\eta \mid {\cal M}\mid^2 = &\left( 
\frac{p\cdot\epsilon^*}{p\cdot k}\right)^2 \left[
N^2 m^2_{\rho}-4 m^2_\pi
+4\left(\frac{p\cdot k}{m_{\rho}}\right)^2\Delta^2
\right]\nonumber\\
&-\epsilon^*\cdot\epsilon
\left[{\bf g}\frac{p\cdot  k}{q\cdot  k}-2+N\Delta
\right]^2
,\label{apf}
\end{eqnarray}
}

\noindent which is free of ${\cal O}(\omega^{-1})$ terms, independently of the MDM value, in accordance with the BK theorem.\\
 An open task is to exploit the relative magnitudes when ${\bf g}\neq 2$ which is expected to be the case due to  structure effects of the vector mesons, and then the possibility of an experimental determination of the MDM. It is clear that an experimental set up is farther difficult than the non-polarized case. Because the properties of the vector mesons are similar to those of the deuteron, except the lifetime,  it could also be used to get a more precise value of its MDM.\\

\section{III. INTERFERENCES OF GAUGE INVARIANT AMPLITUDES}

Another features which are strongly related with the gauge invariant structure of the radiative amplitudes will be shown to appear in  the interferences between the electric charge radiation, namely   ${\cal M}_e$ identified with $A/\omega$ in Eq. (1), and higher order contributions to the amplitude, which can be of diverse origin but always gauge invariant by themselves.\\
Let us start by defining the tensor:
\be
L^\mu \equiv \left( \frac{p^\mu}{p\cdot k} - \frac{q^\mu}{q\cdot k}\right)
\ee

\noindent which satisfies that $L\cdot k=0$ and therefore any term exhibiting a proportionality to it will be gauge invariant. Turns out that the electric charge radiation has this dependence. In what follows we will exploit this fact. In the rest frame of the decaying particle, the square of this tensor is:

\be 
L^2=L \cdot L =-\frac{|\vec{p}|^2}{(p\cdot
k)^2} \sin^2\theta
\ee
where $\theta$ is the angle between the photon three-momentum and 
$\vec{p}$.\\
Let us now summarize the results of the computed interferences for definite processes to exhibit a common behavior and then make a general statement on the interferences base on gauge invariance and the form of the electric charge radiation.\\

{\it Vector meson decays.\\}
Considering the amplitude for the decay  $\rho^- (q,\eta) \rightarrow \pi^- (p) \pi^0 (p') \gamma(k,\epsilon)$ \cite{toledoJG01}. The model-dependent contributions to the term $C$ in Eq. (1) are dominated by those terms coming from intermediate  mesons through the sequence:
\[
\rho^- \rightarrow \pi^- X^0 \rightarrow \pi^-\pi^0\gamma
\]

\noindent where $X^0$ is an isoscalar vector meson ($\omega,\phi$) or a neutral  axial meson ($a_1$). The $\omega$ vector meson is the dominant identified source of model-dependent contributions ${\cal M}_d $. The interference  with the term  ${\cal M}_e$ in the Low's amplitude, is:
\be
\sum_{pol}{\cal M}_d{\cal M}^*_e =-\frac{2eg g_{\omega\pi^0 \gamma}
g_{\rho\omega\pi} k\cdot p (k\cdot q)^2}
 {m^2_\rho + m^2_\pi - m^2_\omega-2q\cdot p} L^2.
\ee

The contribution of the electric quadrupole moment of the $\rho$ meson 
${\cal M}_{Q}$, upon the interference with ${\cal M}_{e}$ is:
\be
\sum_{pol}{\cal M}_{Q}{\cal M}^*_e=(eg)^2\gamma(0)p\cdot k \left\{ 
4p\cdot k + \frac{q\cdot k}{m_\rho^2}(2p -q)\cdot k \right\}L^2
\ee

{\it Production of vector mesons.\\} 
Considering the decay  $\tau^-(q) \rightarrow \rho^-(p,\eta)
\nu(p')\gamma(k,\epsilon)$  \cite{toledoJG01}. The charge radiation term of this amplitude, denoted  again by ${\cal M}_e$, also exhibit the proportionality to the $L^\mu$ tensor. We can  identify the model-dependent contribution:   $\tau^- \rightarrow \pi^- \nu \rightarrow \rho^- \nu \gamma$. The corresponding gauge-invariant  amplitude, ${\cal M}_\pi$ {\it vanishes} identically when interfering with the electric charge: $\sum_{pol} {\cal M}_e {\cal M}^*_{\pi}=0$.\\

Another possible model-dependent contribution arises from   $\tau \rightarrow \nu a_1 \rightarrow \nu \gamma \rho$, whose gauge-invariant amplitude ${\cal M}_{a_1}$ upon {\it interference with the electric charge} radiation becomes (except for coupling constants):
\be
\sum_{pol} Re {\cal M}_{a_1} {\cal M}_e^* = \frac{k\cdot p  k\cdot q L^2}{( 2k\cdot p-m_{a_1}^2+m_\rho^2 )} 
\left( \frac{k\cdot p}{m_\rho^2}-2  \right) \ 
\ee

The contribution of the {\it electric quadrupole moment} to the amplitude,
${\cal M}_Q$, introduced in gauge-invariant form, upon interference is given by:
\[
\sum_{pol}{\cal M}_e{\cal M}^*_Q= 
\]
\be
\gamma(0)q\cdot k \left(-4q.k
+\frac{p.k}{m_{\rho}^2}(2[m_{\tau}^2+m_{\rho}^2-q.k]+p.k) \right) L^2\, 
\ee

Therefore all the contributions are proportional to {\bf $L^2$}. Note that in these definite examples the kind of particles involved are diverse: pseudoscalars, vector mesons and spinors, and therefore the common structure has to have an origin not based on a particular field, but the gauge invariance which is a requirement for any single amplitude that interfere with the electric charge radiation. Thus a general case should be possible to state.\\

{\it General case.\\} 
For a three body radiative decay ${\bf A}^-(p) \rightarrow { \bf B}^-(q) { \bf C}^0(p') {\bf \gamma}(\epsilon, k)$, the total amplitude  can be written as a sum of two terms:

\[
{\cal M} = \epsilon^{\mu} (L_{\mu}M_0 + M_{1,\mu})\ .
\]

The term, $M_0$, is the electric charge amplitude of order $\omega^{-1}$ and explicitly proportional to $L^\mu$, while $M_{1,\mu}$ is a gauge-invariant amplitude ($k\cdot M_1=0$) starting at even order $\omega^0$, and therefore the Burnett-kroll terms should also show this property, besides its vanishing upon the sum of all the contributions of the same order.\\
The interference term between the electric charge and the remaining
amplitude, after summing over polarizations, is given by:

\be
{\cal I}=2Re\left [(-g^{\mu \nu})L_{\mu}\{ \sum_{pols:
A,B,C,}M_0^{*}M_{1,\nu}\}\right ]\ .
\ee

The most general form of the factor within curly brackets  in this
equation can be parametrized in the most general form consistent with Lorentz transformation. The gauge-invariance condition imposed on it to be also proportional to $L_\nu$ with a Lorentz invariant coefficient $a$. Therefore,  after contracting with $L_{\mu}$, ${\cal I}$ becomes:

\[
{\cal I}\sim -a  L^2\ .
\]

 Applications of this result can be manifold. In Figure 1 we just illustrate the possibility of emphasize the MDM dependence in a particular kinematical region to extract its value experimentally.\\

\section{IV. DISCUSSION.}

 We  have computed the polarized transition probability for the process $\rho^+ \rightarrow \pi^+\pi^0 \gamma$. We found that at difference of BK result, the ${\cal O}(\omega^{-1})$ contributions, are not null unless the MDM of the vector meson takes the canonical value ${\bf g}=2$.  No further restriction on its value is required elsewhere and this one is the same that has been found for other authors leads to particular properties of the radiative processes. Our results are important in the sense that qualitatively different behavior arises for the radiation as a consequence of the electromagnetic properties of the vector mesons. On the other hand, the gauge invariance requirements for the amplitude  yields a typical structure to the interferences between the charge radiation and any gauge invariant term. In our analysis we have only exploited the gauge invariance  properties of the contributions and the known form of the radiation off electric charges and therefore a generalization of the result was established. As an example we exploited the  kinematical facility coming from the global factor $L^2$ to explore the importance of the interferences by looking to their relative magnitudes.\\
As a conclusion we want to stress that the commonly shadowed radiation interferences can offer further interesting information about properties of hadronic resonances, via the gauge invariance requirements.\\

\begin{acknowledgments}
 This work was partially supported by Conacyt , M\'exico under grants 41600-A1 and 42026-F and PAPIIT IN112902-3.
\end{acknowledgments}

%para indicar alguna figura, incluya el comentario "%figura_(numero de figura)",
%favor de incluir en la información del artículo los archivos correspondientes a las figuras en .bmp o en .jpg

\begin{figure}
\includegraphics{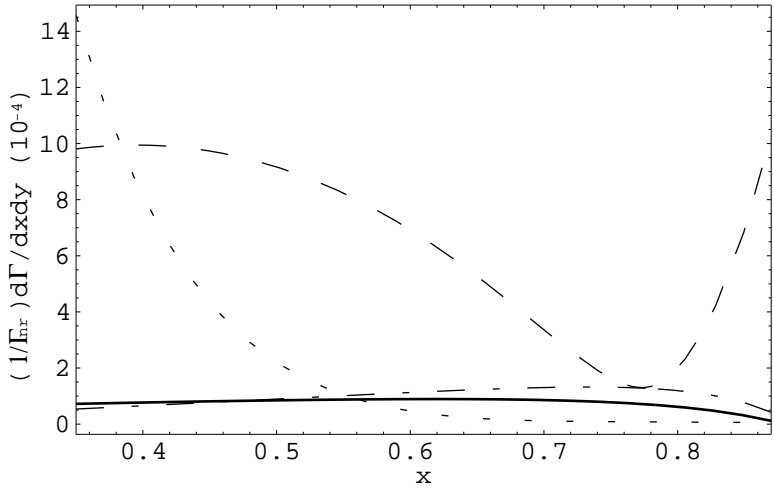}
\caption{ Angular and energy decay distributions of photons in the
process $\rho^+ \rightarrow \pi^+\pi^0 \gamma$, normalized to the
non-radiative rate, as a function of the photon
energy ($x \equiv 2\omega/m_\rho$) for $\theta=10^0$ ($y \equiv \cos\theta$). The short-- and long--dashed lines correspond to the electric charge and the MDM
($g =2$) contributions, respectively. The long-short--dashed and solid lines
correspond to the model-dependent and the electric quadrupole moment
($\gamma(0)=2$) effects, respectively. }
\end{figure}


\begin{thebibliography}{99}
% Incluir Aqui la Bibliografía
\bibitem{pdg}Review of Particle Physics. S. Eidelman et. al.  Phys. Lett. B {\bf 592}(2004).
\bibitem{mackellar} M. B. Hecht and B. H. J. McKellar, Phys. Rev. C{\bf 57},
2638 (1998);F. T. Hawes and M. A. Pichowsky, Phys. Rev. C{\bf 59}, 1743(1999). F. Cardarelli, I. L. Grach. et. al., Phys.~Lett. B{\bf 349}, 393(1995); T. M. Aliev, I. Kanik, M. Savci Phys.Rev. D68 056002(2003).
\bibitem{jackiw}R. Jackiw, Phys. Rev. D{\bf 57}, 2635 (1998)
\bibitem{ferrara}Ferrara, M.~Porrati and V. L. Teledgi, Phys. Rev. D{\bf 46}, 3529 (1992).
\bibitem{bargman}V.~Bargmann, L.~Michel and V.L.~Teledgi, Phys.~Rev.~Lett. {\bf 2}, 433(1959).
\bibitem{bando}M. Bando, T. Kugo and K. Yamawaki,  Nucl. Phys. B {\bf 259} 
 493(1985); Phys. Rept. {\bf 164} 217(1988);
\bibitem{burnett}T. H. Burnett and N. M. Kroll, Phys. Rev. Lett. {\bf 20}, 86(1968).
\bibitem{toledo9700} G. L\'opez Castro and G. Toledo S\'anchez, Phys. Rev. D{\bf56}, 4408 (1997); Phys.~Rev.~D {\bf 60}, 053004 (1999);  Phys.~Rev.~D {\bf 61}, 033007 (2000)
\bibitem{low}F. E. Low, Phys. Rev. {\bf 110}, 974 (1958).
\bibitem{nieves}Hagiwara et al. Nucl. Phys. B {\bf 282}, 253(1987); J. F. Nieves and B. P. Palash, Phys. Rev. D{\bf 55}, 3118(1997).
\bibitem{toledoPRD02}G. Toledo S\'anchez PRD {\bf 66} 097301 (2002)
\bibitem{samuel}R. Mikaelian, M. A. Samuel and D.Sahdev, Phys. Rev. Lett.{\bf 43}, 746 (1979); F. Boudjema, C. Hamzaoui, M. A. Samuel  and J.
Woodside, Phys. Rev. Lett. {\bf 63}, 1906 (1989).
\bibitem{toledoJG01}G. L\'opez Castro and G. Toledo S\'anchez
J. of Phys. G {\bf 27} 2203(2001).



\end{thebibliography}
\end{document}